\documentstyle[prl,twocolumn,aps]{revtex}
\input{epsf}

\newcommand{\bq}{\begin{equation}}
\newcommand{\ee}{\end{equation}}
\newcommand{\fr}[2]{\frac{#1}{#2}}
\newcommand{\eps}{\varepsilon}

\begin{document}
\draft

\title{Vortices in a cylinder: Localization after depinning}

\author{P.G.Silvestrov}

\address{Budker Institute of Nuclear Physics, 630090
Novosibirsk, Russia}

\maketitle

\begin{abstract}

Edge effects in the depinned phase of flux lines in hollow
superconducting cylinder with columnar defects and electric
current along the cylinder are investigated. Far from the ends
of the cylinder vortices are distributed almost uniformly
(delocalized). Nevertheless, near the edges these free vortices
come closer together and form well resolved dense bunches. A
semiclassical picture of this localization after depinning is
described. For a large number of vortices their density
$\rho(x)$ has square root singularity at the border of the bunch
($\rho(x)$ is semicircle in the simplest case). However, by
tuning the strength of current, the various singular regimes for
$\rho(x)$ may be reached. Remarkably, this singular behaviour
reproduces the phase transitions discussed during the past
decade within the random matrix regularization of $2d$-Gravity.

\end{abstract}

\pacs{PACS numbers: 74.60.Ge, 72.15.Rn, 05.45.+b }

Statistics of ensembles of flux lines in high-$T_c$
superconductors has been a subject of numerous experimental and
theoretical investigations (see for review Ref.~\cite{Larkin}).
An important modification of the superconductor is achieved by
creation of artificial disorder in the form of columnar pins
produced by heavy ion irradiation. The vortex distribution in
the cylindrical superconductors with columnar defects and
longitudinal electric current has attracted a renewed interest
recently~\cite{Hatano}. The current creates a transverse
magnetic field, which attempts to wind vortices around the
cylinder. For low current the vortices are trapped by the
fluctuations in the density of defects and do not curl
({transverse Meissner effect} \cite{Vinokur}). With the increase
of current the transition to depinned phase takes place. The
mapping of this transition onto the delocalization transition in
$1d$ non-Hermitean quantum mechanics \cite{Hatano} has caused an
immediate and wide interest \cite{Efetov}. However, as it was
pointed out in Ref.~\cite{Silv}, even after the transition to
complex spectrum eigenfunctions of corresponding non-Hermitean
Hamiltonian still exhibit the features of both localized and
delocalized states. Physical consequences of this "localization
after depinning" for vortices in the thin-shelled cylinder will
be the subject of this paper. Our main prediction is illustrated
by the Fig.~1.  The depinned fraction of vortices is practically
uniformly distributed over the surface far from the ends of
cylinder.  However, while approaching the ends, vortices come
closer and form well resolved localized bunches. The density of
vortices in this bunch at the edge coincides with the density of
eigenvalues for the ensembles (Gaussian or non-Gaussian) of
orthogonal random matrices. The phase transitions emerging while
tuning the parameters in these random matrix models were a
subject of intensive investigation in the past decade in the
context of problems of $2d$-Gravity \cite{Ginsparg}. As we show
in this paper, experimental investigation of vortices in
cylindrical samples may open the way for the direct observation
of such transitions in real low-dimensional systems.

The classical energy for a flux line in the cylinder may be
written as an action of equivalent particle \cite{units}
\bq\label{action}
S=\int_0^{L_\tau} \left[ \fr{1}{2}\left( \fr{dx}{d\tau}
\right)^2 
-h\fr{dx}{d\tau}  +V(x)\right] d\tau \ .
\ee
The length and period of the cylinder are $L_\tau$ and $l$,
$x+l\equiv x$. The potential $V(x)$ accounts for the interaction
of vortex with the columnar defects and $h$ is proportional to
the longitudinal current. All our results are valid for random
$V(x)$. However, in order to enhance the effect it is better to
prepare the sample with smooth and inhomogeneous at the scale
$\Delta x\sim l$ density of columnar pins. The partition
function for classical vortex now takes the form of quantum
evolution operator in imaginary time
\bq\label{Green}
Z_{L_\tau}(x_2,x_1)= \int {\cal D}x e^{-\beta S}
=\langle x_2| e^{-\beta HL_{\tau}}
|x_1\rangle \ ,
\ee
where $\beta=1/T$ is the inverse temperature and we have
introduced the non-Hermitean Hamiltonian 
\bq\label{Ham}
H= -\fr{1}{2}\left( \fr{1}{\beta}\fr{\partial}{\partial x}
-h\right)^2 + V(x) \ .
\ee
Thus the electric current in the cylinder acts like an imaginary
vector potential $ih$.  The path integral in Eq.~(\ref{Green})
includes strings with given end points $x_1$ and $x_2$. In order
to find the partition function for vortex with free ends one has
to integrate over $x_{1,2}$ \cite{Hatano}. The evolution
operator Eq.~(\ref{Green}) may be written in the form
\bq\label{GSum}
Z_{L_\tau}(x_2,x_1)=\sum_i \psi^R_i(x_2) \psi^L_i(x_1)
e^{-\beta\eps_i L_\tau} \ 
, 
\ee
where the left- and right- eigenvectors are defined via $H\psi^R
=\eps\psi^R$ and $H^T\psi^L =\eps\psi^L$. The solutions are
normalized via $\int_0^l \psi^L_i \psi^R_j dx = \delta_{ij}$. In
this paper we will consider only the case $h^{-1}\ll l\ll
{L_\tau}$. In particular this means that only the ground state
contribution survives in the sum in Eq.~(\ref{GSum}). 

Interacting vortices in this approach are equivalent to the
interacting bosons. Moreover, we will consider only the case of
strong $\delta$-function repulsion of vortices(bosons).

Interesting for us physical information is contained in the slow
harmonics of the potential $V(x)$. Therefore, in this paper we
will restrict our attention on the case $dV/dx\ll V$. Also we
will consider the case of strong depinning current $h^2 \gg V$.
\bq\label{WKB}
\psi =\fr{\exp(-\sigma)}{\sqrt{l}} \ , \
\sigma= \beta \int_0^x (\sqrt{2(V-\eps)} -h)dx' + ... \ .
\ee
The corrections to $\sigma$ may be also easily found. 
The quantization condition, which allows one
to find the set of complex eigenvalues $\eps_n$, is $\sigma(l)-
\sigma(0)=2\pi n i$. It will be enough for us to consider only
low lying excitations of the Hamiltonian Eq.~(\ref{Ham}). In
this case  
\begin{eqnarray}\label{En}
&&\eps_n =-(ikT-h)^2/2 +\langle V\rangle \ \ , \ \ k=2\pi n/l \ \
, \\ 
&& \sigma_n(x) = - ikx + \int_0^x \fr{V(x')- \langle V\rangle
}{hT} 
dx' \ . \nonumber
\end{eqnarray}
Here $\langle V\rangle =l^{-1}\int_0^l Vdx$. The most exciting
in this result is that $\sigma(x)$ acquires the nontrivial real
part in the presence of columnar field $V(x)$ (in general $|Re
\sigma|\gg 1$ if $l\gg h/V$). This means that even
after the depinning transition eigenvectors of the non-Hermitean
Hamiltonian Eq.~(\ref{Ham}) still remain strongly
(exponentially!) localized. The localization described by the
Eqs.~(\ref{WKB},\ref{En}) has almost nothing in common with
usual localization at $h=0$. To see this consider the
simple example of very wide square-well potential: $V=U_0$ for
$0<x<l/2$ and $V=-U_0$ for $-l/2<x<0$. Instead of almost
extended ground state in the absence of transverse field
$\psi(h=0)\sim \sin(2\pi x/l)$ at $-l/2<x<0$ one finds via the
Eq.~(\ref{En}) the clearly localized even for $h^2\gg U_0$
eigenvector $\psi_0^R \sim \exp(-U_0|x|/h)$ \cite{footnote}.
Moreover, we see from this example that the imaginary vector
potential first creates the localized states (maximum of
localization is reached at $h^2\sim V-\langle V\rangle$).
However, with further increase of $h$ the localization length
grows up again. All eigenstates described by the Eq.~(\ref{En})
are localized near the minimum of $\sigma_0(x)$. This feature of
"non-Hermitean" localization is also in sharp contrast with the
usual Anderson case, where eigenstates with close energy
strongly repel in the coordinate space.

The Eqs.~(\ref{WKB},\ref{En}) describe localization of
eigenvectors by the long wave-length harmonics of potential
$V(x)$ in strong imaginary vector potential. In the case of
random disorder, however, a part of eigenfunctions (whose with
the localization length $\xi <h^{-1}$) may be localized in the
usual Anderson sense on the rare local fluctuations of $V(x)$
\cite{Hatano}. In this case, if there are only few vortices in
the sample, they will be trapped by these true localized states.
However, with increase of the number of fluxes, the fraction of
depinned vortices emerges. Corresponding wave functions are
described again by the analog of Eqs.~(\ref{WKB},\ref{En}). Just
this situation is shown on the Fig.~1.

The localization described by the WKB formula Eq.~(\ref{WKB})
should have a simple classical explanation. Indeed, the action
for the solution of classical equation of motion for the string
connecting points $x_1$ and $x_2$ with $n$ windings over the
cylinder may be shown to have a form
\bq\label{Sn}
S_{cl}^{(n)}= \eps L_{\tau} +\int_{x_1}^{nl+x_2}(\sqrt{2(V-\eps
)} -h)dx 
\ \ ,
\ee
where the classical energy $\eps$ should be found from
$\int_{x_1}^{nl+x_2} dx/\sqrt{2(V-\eps)}= L_{\tau}$. Note that
we use the reversed sign of the kinetic energy
$\eps=-\dot{x}^2/2+V$ compared to usual classical definition. In
order to find the most important classical configuration we have
to determine the minimum of $S^{(n)}$ with respect to $n$. For
$n\gg 1$ one has for $\Delta S_{cl}=S_{cl}^{(n+1)}-S_{cl}^{(n)}$
\bq\label{DSn}
\Delta S_{cl}= \delta \eps\left( L_{\tau} - L_{\tau}\right) 
+ \int_{0}^{l}(\sqrt{2(V-\eps)} -h)dx \ ,
\ee
where $\delta \eps$ is the energy difference for two paths.
Thus, the condition of extremum $dS^{(n)}/dn=0$ coincides with
the semi-classical quantization condition for the ground state
of Hamiltonian Eq.~(\ref{Ham}). Consequently, the contribution
of smallest $S_{cl}^{(n)}$ into the functional integral
Eq.~(\ref{Green}) is enough to reproduce the WKB result
Eqs.~(\ref{GSum},\ref{WKB}) in the limit of large $L_\tau$. The
classical consideration allows also to find the typical tilt of
the vortex trajectory $\langle \dot{x} \rangle = h+O(V^2/h^3)$.

The Hamiltonian Eq.~(\ref{Ham}) and its transpose are related
via $H^T(h)=H(-h)$ \cite{Hatano}. This simple equality allows
one to introduce the conserving current for left- and right-
eigenvectors with the same energy $J={\psi^{L}}'\psi^R-
\psi^L{\psi^R}' +2h\psi^L\psi^R \ ; \ J'\equiv dJ/dx =0$. For
small $h$ and random $V(x)$ in the thermodynamic limit
$l\rightarrow \infty$ one has $J=0$ for all low (real)energy
states. Above some critical value $Re\eps=\eps_c$ the current
become non-zero and the energy acquires nontrivial imaginary
part.  This appearance of conserving current $J\ne 0$ was
usually considered as an indication of existence of
delocalization transition in $1d$ non-Hermitean quantum
mechanics. However, formally the existence of current shows the
absence of exponential localization only for quantities bilinear
in $\psi^L$ and $\psi^R$. Individually $\psi^L$ and $\psi^R$
still may be localized. Indeed, for the WKB wave functions
Eq.~(\ref{WKB}) one finds $\psi^L(x)\psi^R(x)=l^{-1}$, i.e.
exponential growth of $\psi^R$ is compensated by the same
decrease of $\psi^L$.  The probability to find vortex at the
point $x$ on the transverse slice $\tau$ of the cylinder due to
the Eq.~(\ref{Green}) has the form $\rho(x,\tau) = Z^{-1}
\langle \Psi^f| e^{-\beta(L_{\tau}-\tau)H} |x\rangle \langle
x|e^{-\beta\tau H}|\Psi^i\rangle$. For large $\tau,L_{\tau}-\tau
\sim L_{\tau}$ one finds $\rho(x,\tau) \sim \psi_0^R
(x)\psi_0^L(x)$. We see that the effect of localization is
washed out inside the cylinder. However \cite{Hatano,Tauber}, at
the top of cylinder $\rho^R(x)=\rho(x,L_{\tau}) \sim \psi_0^R
(x)$ and at the bottom $\rho^L(x)=\rho(x,0) \sim \psi_0^L(x)$
the vortex is strongly localized.

As we have told, the ensemble of vortices should be treated as
the system of interacting bosons with single particle
Hamiltonian Eq.~(\ref{Ham}). At least for low density of fluxes,
the inter-boson interaction is equivalent to short-range
repulsion $U=U_0\delta(x-x')$. For many vortices the
single-particle wave functions $\psi^{R,L}_i$ and energy
$\eps_i$ should be replaced by  the many-particle ones
$\Psi^{R,L}_i$ and the total energy $E$.  The ground state for
impenetrable bosons is found by the analogy with $N$-fermion
solution \cite{Girardeu}
\bq\label{PsiN}
\Psi^R=\prod_{i<j} \left( \prod_l sign(x_i
-x_j+2\pi l)\right) \fr{det|\psi_n^R(x_m)|}{\sqrt{N!}} \ .
\ee
Here $\psi_n^R$ are the eigenvectors of the single particle
Hamiltonian Eq.~(\ref{Ham}) with lowest $Re\eps_n$. For odd
$N$ the eigenvectors should be the usual periodic
$\psi_n^R(x+l)=\psi_n^R(x)$. However, for even $N$ one should
use the antiperiodic boundary conditions $\psi_n^R(x+l)=
-\psi_n^R(x)$. In both cases the total ground state energy
$E=\sum\eps_n$ entering the analog of the Eq.~(\ref{GSum}) is
real. Also for the even and odd $N$ and smooth $V(x)$
the $\psi_n^R(x)$ are well described by the
Eqs.~(\ref{WKB},\ref{En}). Within this approximation the wave
function is further simplified
\bq\label{PsiNWKB}
\Psi^R=\fr{1}{\sqrt{N!}}\prod_m
\fr{e^{-\sigma_0(x_m)}}{\sqrt{l}} 
\prod_{i<j} 2 \left| \sin \left(\fr{\pi(x_i-x_j)}{l}\right)
\right| \ .
\ee
Furthermore, in the most interesting case, when the function
$\exp(-\sigma_0(x_m))$ has a narrow ($\Delta x \ll l$) peak one
may replace $| \sin ({\pi(x_i-x_j)}/{l}) |$ by $\pi|x_i-x_j|/l$.
This means that the distribution of vortices described by the
$\Psi^R$ coincides formally with the distribution of eigenvalues
of $N\times N$ Orthogonal Random Matrix Ensemble with the weight
function $\exp(-Tr \sigma_0(M))$. The density of vortices
$\rho(x)$ at the end of the cylinder is found after integration
of the $\Psi^R$ over the all $x_i$ except for one. The
saddle-point method (large $N$ approximation) for calculation of
such integrals was developed many years ago in
Ref.~\cite{Itzykson}.  In particular, in the most general case
of quadratic minimum of $\sigma_0$ one has
\bq\label{circle}
\sigma_0 =\fr{\alpha(x-x_0)^2}{2hT} \ ; \ \rho=\fr{\alpha}{\pi
hT} 
\sqrt{\fr{2hT}{\alpha}N-(x-x_0)^2} \ .
\ee
Here we have shown explicitly the dependence of $\sigma_0$ on
the transverse field $h$ and temperature $T$. The coefficient
$\alpha$ is determined only by the potential $V(x)$.
With the increase of $N$ the anharmonic contributions to
$\sigma_0$ should be also taken into account. For example one
may write (we put $x_0=0$)
\bq\label{general}
\sigma_0 =\fr{1}{hT} \fr{\alpha x^2}{2} W(x/\lambda) \ , \
W(0)=1 \ .
\ee
Now $\alpha$ contains the information about the strength of
interaction $V(x)$, while $\lambda$ is the characteristic length
for its variation. The vortices at the edge of the cylinder in
this case also form a smooth dense bunch  with square root
$\rho(x)$ at the border (see Fig.~2a). The width of the bunch is
$x_c\sim \sqrt{NhT/\alpha}$ and the anharmonic contributions
became important starting from $N\sim N_c=\alpha\lambda^2/hT$.
The new phenomena may take place if the function $\sigma_0(x)$
has more than one minimum. Some variants of a peculiar
behavior of $\rho(x)$ in this case are illustrated by Figs.~2b
and 2c. With increase of $N$ in the case of two minimums the
second small stable bunch of vortices is born 
at some value $N_1$. These two bunches join
together at the second critical value $N_2$. The moment of
consolidation of two bunches into one is shown on the Fig.~2b.
The density of states close to the transition in the vicinity of
meeting point $x_0$ is
\begin{eqnarray}\label{meet}
\rho=a(x-x_0)\sqrt{(x-x_0)^2+\Delta  [h-h_c]} \ \ 
&{\mbox {for}}& \ \  
h<h_c \, , \nonumber \\
\rho=a((x-x_0)^2+\Delta  [h-h_c]/2) \ \  &{\mbox {for}}& \ \ 
h>h_c \, , %\nonumber
\end{eqnarray}
where $a$, $\Delta$, $x_0$ and $h_c$ vary smoothly with the
change of $N$, $T$, or $h$. The number of vortices which may be
kept in equilibrium in each well may be regulated by the
external parameter $h$. This feature of the vortex distribution
opens the way to create the metastable configurations like one
shown on the Fig.~2c. For $h<h_c^{*}$ the vortices are confined
(during exponentially long time) in the deepest well. Above
$h_c^{*}$ the fast decay into second well takes place. The
critical configuration shown on the Fig.~2c is characterized by
the novel singular behaviour of the density at the border
$\rho(x)\sim (x-x_0)^{3/2}$. Just this type of critical
behaviour corresponds to the continuous limit in the Random
Matrix regularization of $2d$-Gravity.

The catastrophic change in vortex distribution at the transitions
should also change the thermodynamic characteristics of the
ensemble of flux lines. After integration over the positions of
the ends of vortices one finds the partition function and the
Free energy
\bq\label{Free}
F=-T\ln(Z)=F_{L_\tau}+F_{e}^R+F_{e}^L \ .
\ee
The contribution proportional to the total volume $F_{L_\tau}$
in our simple case has the form
\bq\label{FL}
F_{L_\tau}=NL_\tau\left\{ -{h^2}/{2} +\langle V\rangle +
{(\pi NT/l)^2}/{6}+f_0 \right\} \ .
\ee
Here $f_0(T)$ accounts for the short wave-length fluctuations of
string and does not depend on $h$, 
$V(x)$ and $N$. The more interesting for us
are the edge contributions $F_{e}^R$ and $F_{e}^L$. For large
$N$ and the quadratic $\sigma_0$ 
\bq\label{Fedge}
F_{e}^R = N^2 T\left\{ \fr{3}{8}+ \fr{1}{4}\ln\left(
\fr{\alpha^R l^2}{4\pi^2NhT} \right) \right\} + Ng_0 \ .
\ee
Here again $g_0$ depends on the details of regularization of
the functional integral Eq.~(\ref{Green}). With the increase of
$N$ (or $h,T$) the edge Free energy changes smoothly until one
meets one of the singular points considered above. The first
possible singularity is the birth of new small bunch. The
corresponding correction at $N>N_1$ is
\bq\label{birth}
\Delta F_{e}= -AT(N-N_1)^2 \ .
\ee
Here $A=A(N,hT)>0$. As we have learned from the 
$2d$-Gravity, this correction is purely nonperturbative, i.e. it
could not be related with the analytic behaviour of $F_{e}$
below the singularity. Another two kinds of singular corrections
associated with the confluence of two bunches ($h_c$) and decay
of metastable bunch ($h_c^*$) lead to
\bq\label{death}
\fr{\Delta F_{e}}{N^2T}\sim (h_c-h)^{3/2} \ \ {\mbox {and}} \ \
\fr{\Delta F^*_{e}}{N^2T}\sim (h^*_c-h)^{5/2} \ .
\ee
The constants $h_c$ and $h^*_c$ are the functions of $N$ and
$T$. Depending on the concrete way of realization of physical
experiment one may write instead of the Eq.~(\ref{death}), for
example, $\Delta F\sim (T_c-T)^{3/2}$ or $\Delta F\sim
(N_2-N)^{3/2}$.

In summary, the edge effects in vortex distribution in
superconducting cylinders may provide us with the variety of new
phenomena with clear experimental signature (see again the
Figs.~1,2). Among them are the strong localization of vortices
at the end of cylinder and various critical regimes for this
localization available by tuning of the longitudinal current.
Technically these effects arose due to the peculiar features of
localization in non-Hermitean quantum mechanical Hamiltonian
(\ref{Ham}). From the pure theoretical point of view, the most
exciting is the correspondence between distribution of flux
lines at the end of cylinder and distribution of eigenvalues of
the ensembles of random matrices.  The ensemble of fluxes turns
out to be the almost unique example of the system, where not
only local (correlations of close levels etc.), but also global
features of random-matrix spectrum are of 100\% importance. For
example, the phase transitions in non-Gaussian Matrix ensembles
have been a subject of enormous activity in last 10 years within
the context of $2d$-Gravity \cite{Ginsparg}. However, to the
best of my knowledge, in this paper the first proposal is
presented of a real physical experiment where such a singular
behaviour may be observed.

Author is thankful to N.~Hatano and A.~Zee whose comments on the
paper Ref.~\cite{Silv} have initiated this work. Valuable
discussions with M.~V.~Mostovoy, D.~V.~Savin, V.~V.~Sokolov,
O.~P.~Sushkov and A.~S.~Yelkhovsky are greatly acknowledged.
The work was supported by RFBR, grant 98-02-17905.

\vspace{-.8cm}
\begin{figure}[t]
\epsfxsize=8.5cm%8cm
\epsffile{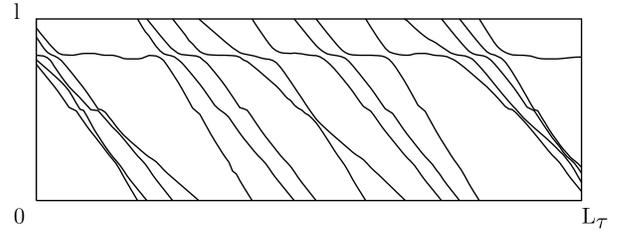}
\vglue 0.2cm
\medskip
\caption{The vortex distribution on
the cylinder ($l\equiv 0$). The
Hamiltonian $H$ has one eigenstate localized in the usual sense
due to a strong local fluctuation of $V(x)$. In the typical
transverse section far from the ends of the cylinder one vortex
is trapped by this localized level, while others are almost
uniformly distributed. However, near the ends
these free vortices come closer together and form well resolved
dense bunches.
}
\end{figure}

\vspace{-.8cm}
\begin{figure}[t]
\epsfxsize=8.5cm%8cm
\epsffile{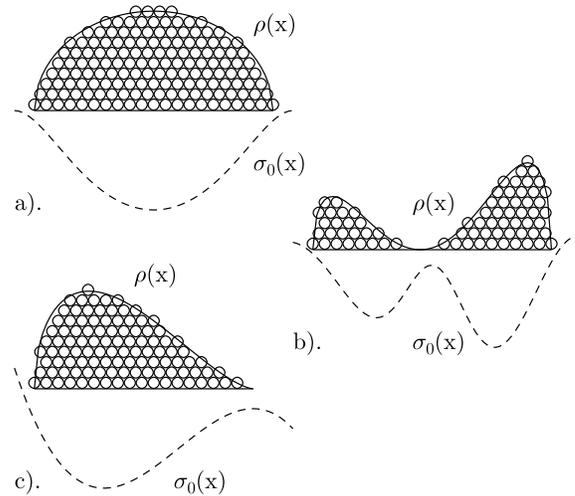}
\vglue 0.2cm
\medskip
\caption{Examples of distribution of vortices at the end of
cylinder: a). Semicircle in a single well b). Confluence of two
bunches in a double well c). Opening of a decay of metastable
bunch. 
}
\end{figure}


\begin{thebibliography}{99}
\vspace{-1cm}

\bibitem{Larkin} {\it Phenomenology and Applications of High
Temperature Superconductors}, edited by K.~Beddel {\it et al.}
(Addison-Wesley, New York, 1991); G.~Blatter {\it et al.},
Rev.~Mod.~Phys, {\bf 66} (1994) 1125.

\bibitem{Hatano} N.~Hatano and D.~R.~Nelson, Phys.~Rev.~Lett.
{\bf 77}, 570 (1996); {\it ibid},
Phys.~Rev.~{\bf B 56},8651 (1997)

\bibitem{Vinokur} D.~R.~Nelson and V.~Vinokur, Phys.~Rev.~{\bf
B~48},13060 (1993).  

\bibitem{Efetov} K.~B.~Efetov, Phys.~Rev.~Lett.  {\bf 79}, 491
(1997), {\it ibid}, Phys.~Rev.~{\bf B 56}, 9630 (1997);
J.~Feinberg, A.~Zee, cond-mat/9706218, cond-mat/9710040;
P.~W.~Brouwer, P.~G.~Silvestrov, and C.~W.~J.~Beenakker,
Phys.~Rev.~{\bf B 56}, R4333 (1997); R.~Janik, M.~A.~Nowak,
G.~Papp, and I.~Zahed, cond-mat/9705098; I.~Y.~Goldsheid and
B.~A.~Khoruzhenko, cond-mat/9707230; E.~Brezin, A.~Zee,
cond-mat/9708029; D.~R.~Nelson and N.~M.~Shnerb,
cond-mat/9708071, cond-mat/9801111.

\bibitem{Silv} P.~G.~Silvestrov, cond-mat/9802219.

\bibitem{Ginsparg} See for review e.g., P.~Di~Francesco,
P.~Ginsparg, J.~Zinn-Justin, Phys.~Rep.  {\bf 254}, 1 (1995).

\bibitem{units} For transition to physical units see e.g.,
Refs.~\cite{Hatano,Vinokur}. 

\bibitem{Tauber} U.~C.~T\"{a}uber and D.~R.~Nelson, Phys.~Rep.
{\bf 289}, 157 (1997). 

\bibitem{footnote} The more accurate solution, without a cusp at
$x=0$, is: $\psi = (1-U_0/h^2)\exp(-U_0 x /h)$ for $x>0$ and
$\psi = \exp(U_0 x /h)- U_0/h^2\exp(2hx)$ for $x<0$.

\bibitem{Girardeu} M.~Girardeu, J.~Math.~Phys. {\bf 1}, 516
(1960); E.~Lieb and W.~Liniger, Phys.~Rev. {\bf 130}, 1605
(1963); A.~Lenard,  J.~Math.~Phys. {\bf 5}, 930 (1964).

\bibitem{Itzykson} E.~Brezin, C.~Itzykson, G.~Parisi, and
J.~B.~Zuber, Comm. Math. Phys. {\bf 58}, 35 (1978). 

\end{thebibliography}
\end{document}